\documentclass[]{emulateapj}
\usepackage{apjfonts}
\usepackage{natbib}
\usepackage{color}
\usepackage{graphicx}
\shorttitle{ }
\shortauthors{ }

\begin{document}

\title{Slow quenching of star formation in OMEGAWINGS clusters: galaxies in transition in the local universe
}

\author{A. Paccagnella\altaffilmark{1,2}}
\author{B. Vulcani\altaffilmark{3}}
\author{B. M. Poggianti\altaffilmark{2}}
\author{A. Moretti\altaffilmark{1,2}}
\author{J. Fritz\altaffilmark{4}}
\author{M. Gullieuszik\altaffilmark{2}}
\author{W. Couch\altaffilmark{5}}
\author{D. Bettoni\altaffilmark{2}}
\author{A. Cava\altaffilmark{6}}
\author{M. D'Onofrio\altaffilmark{1,2}}
\author{G. Fasano\altaffilmark{2}}
\affil{\altaffilmark{1}Department of Physics and Astronomy, University of Padova, Vicolo Osservatorio 3, 35122 Padova, Italy}
\affil{\altaffilmark{2}INAF - Astronomical Observatory of Padova, 35122 Padova, Italy}
\affil{\altaffilmark{3}Kavli Institute for the Physics and Mathematics of the Universe (WPI), Todai Institutes for Advanced Study, the University of Tokyo, Kashiwa, Japan}
\affil{\altaffilmark{4}Instituto de Radioastronom\'\i a y Astrof\'\i sica, IRyA, UNAM, Campus Morelia, A.P. 3-72, C.P. 58089, Mexico}
\affil{\altaffilmark{5}Australian Astronomical Observatory, PO Box 915, North Ryde, NSW 1670 Australia}
\affil{\altaffilmark{6}Observatoire de Gen\`eve, Universit\`e de Gen\`eve, 51 Ch. des Maillettes, 1290 Versoix, Switzerland}

\begin{abstract}
The star formation quenching depends on environment, but a full understanding of what mechanisms drive it is still missing.
Exploiting a sample of galaxies with masses $M_\ast>10^{9.8}M_\odot$, drawn from the WIde-field Nearby Galaxy-cluster Survey (WINGS) and its recent extension OMEGAWINGS, we investigate the star formation rate (SFR) as a function of stellar mass (M$_*$) in galaxy clusters at $0.04<z<0.07$. We use non-member galaxies at 0.02$<$z$<$0.09 as field control sample. Overall, we find agreement between the SFR-M$_*$ relation in the two environments, but detect a population of cluster galaxies with reduced SFRs which is rare in the field. 
These {\it transition} galaxies are mainly found within the cluster virial radius ($R_{200}$) but they impact on the SFR-M$_*$ relation only within 0.6R$_{200}$. 
The ratio of transition to PSF galaxies strongly depends on environment, being larger than 0.6 within 0.3R$_{200}$ and rapidly decreasing with distance, while it is almost flat with $M_*$. As galaxies move downward from the SFR-M$_*$ main sequence, they become redder and present older luminosity and mass weighted ages. These trends, together with the analysis of the star formation histories, suggest that transition galaxies have had a reduced SFR for the past 2-5 Gyr. Our results are consistent with the hypothesis that the interaction of galaxies with the intracluster medium via strangulation causes a gradual shut down of star formation, giving birth to an evolved population of galaxies in transition from being star forming to becoming passive.

\end{abstract}


\section{Introduction}
The distribution of many galaxy properties, including color, morphology and star formation rate (SFR), is bimodal, reflecting the existence of two broad types of galaxies: red, old, quiescent early-type galaxies and dust-reddened or blue, late-type disk galaxies with ongoing star formation. 
This bimodality is related to the variations in the cold gas content, which lead to different levels of star formation and eventually to quenching. Whether the reasons for galaxies getting quenched are internal or external is of critical importance.

One way to shed light on the processes responsible for quenching is to compare the SFR of galaxies of a given stellar mass (M$_*$) in different environments. As shown by \cite{2014ApJ...785L..36A}, a clearer picture can be obtained considering separately relationships with bulge- and disk-masses.

The  correlation found between SFR and the total M$_*$ \citep[main sequence, MS,][]{2004MNRAS.351.1151B,2007ApJ...660L..43N,2007A&A...468...33E,2007ApJ...670..156D} has been interpreted as the result of the balancing of inflows of cosmological gas and outflows due to the feedback \citep{2010ApJ...718.1001B,2013ApJ...772..119L}, thus being sensitive to any physical mechanism affecting the amount of gas available for star formation.
The star formation quenching seems to be stronger in clusters, which have a higher fraction of early-types and a lower fraction of late-type galaxies than the field \citep[e.g., ][]{1980ApJ...236..351D, 1999ApJ...518..576P, 2009ApJ...693.1840B}, suggesting that they are extremely effective in cutting off the galaxy's ability to form stars.

Examples of external processes acting in high-density regions include ram-pressure stripping \citep{1972ApJ...176....1G}, high speed galaxy encounters \citep[galaxy harassment; ][]{1996Natur.379..613M}, galaxy-galaxy mergers \citep{1994ApJ...425L..13M}, and removal of the warm and hot halo gas \citep[strangulation; ][]{1980ApJ...237..692L, 2000ApJ...540..113B}.

The most direct way to isolate the effects of the environment is to study the trends in the galaxy population mix in clusters, where most of the aforementioned processes are strongly active, and compare them with the field. Furthermore, it is  useful to analyze the radial distribution of member galaxies and how their properties change within the clusters, since the cluster-centric distance traces the cluster density profile and is an approximate time-scale for the star formation quenching \citep{2004MNRAS.352L...1G}.

While so far the SFR-$M_*$ relation has been largely investigated in the field, only few studies focused on clusters. 
Using local galaxy field samples, \citet{2010ApJ...721..193P} and \citet{2012MNRAS.423.3679W} showed that the local  density mostly changes the fraction of passive galaxies but has little effect on the SFR-M$_*$ relation.
Focusing on one cluster at the time, also \cite{2013ApJ...773...86T} and \cite{2014ApJ...794...31T} failed to identify any difference in the SFR distribution of cluster and field galaxies and did not detect any population of galaxies departing from the MS. This implies that the quenching of star formation  is a relatively fast transition. 
On the other hand, \citet{2010MNRAS.404.1231V} ($z<0.1$) and \citet{2013ApJ...775..126H} ($0.15<z<0.3$)  found evidence that galaxies in clusters have lower SFRs than galaxies of similar mass in the field.
A population of galaxies with reduced SFR at given mass was discovered in clusters by \citet{2009ApJ...705L..67P} at $z \sim 0.8$ and  \citet{2010ApJ...710L...1V}  at $0.4<z<0.8$.

What is still missing is a detailed study of the local cluster population in a statistically meaningful sample. This is now possible thanks to the OMEGAWINGS survey, which covers 46 clusters at $0.04<z<0.07$, each over a field of 1deg$^2$, allowing us to study the properties of $\sim 5000$ galaxies in an homogeneous sample.

Our analysis  considers all galaxies together, regardless of their morphology. A detailed study as a function  of morphologies is underway (Calvi et al. in prep., Paccagnella et al. in prep.).

We adopt a \citet{1955ApJ...121..161S} initial mass function in the mass range 0.15-120 M$_{\odot}$. The cosmological constants assumed are $\Omega_m=0.3$, $\Omega_{\Lambda}=0.7$ and H$_0=70$ km s$^{-1}$ Mpc$^{-1}$.

\section{Data-set}
The galaxy sample is extracted from the WIde-field Nearby Galaxy-cluster Survey (WINGS) \citep{2006A&A...445..805F, 2014A&A...564A.138M}, that includes 76 X-ray selected clusters of galaxies with $0.04<z<0.07$, and from the recent extension of this project, OMEGAWINGS, that quadruples the area covered by the optical imaging for 46 clusters (\citealt{2015A&A...581A..41G}, Moretti et al. in prep.). The cluster sample covers a wide range of velocity dispersion ($\sigma_{cl}\sim$ 500-1300 km/s) and X-ray luminosity (L$_X\sim$ 0.2-5 $\times 10^{44}$ erg/s).

The spectroscopic targets were selected based on B and V photometry. The spectroscopic catalogs have been corrected for both geometrical and magnitude incompleteness, using the ratio of number of spectra yielding a redshift to the total number of galaxies in the parent photometric catalog, calculated both as a function of V magnitude and radial projected distance from the brightest cluster galaxy (BCG).
In this work, 31 clusters with a spectroscopic completeness higher than 50$\%$ are considered.

A galaxy is considered a cluster member if its redshift lies within $\pm3\sigma_{cl}$ from the cluster mean redshift \citep[see ][] {2009A&A...495..707C}. 
The virial radius R$_{200}$ is computed from $\sigma_{cl}$ following \citet{2006ApJ...642..188P} and used to scale the cluster-centric distances of member galaxies. The field control sample is extracted from the non-member galaxies in the redshift range $0.02<z<0.09$.

We derive SFRs, star formation histories (SFHs), $M_*$, luminosity- and mass- 
weighted ages (LWA and MWA, respectively) by fitting the spectra with the spectrophotometric model fully described in \citet{2007A&A...470..137F, 2011A&A...526A..45F,2014A&A...566A..32F}. All the main spectrophotometric features (i.e. the continuum flux and shape, the equivalent widths of emission and absorption lines) are reproduced by summing the theoretical spectra of simple stellar populations of 12 different ages (from $3\times 10^6$ to approximately $14\times 10^9$ years).

Stellar masses used are masses locked into stars, including both those that are still in the nuclear-burning phase, and remnants such as white dwarfs, neutron stars and stellar black holes.

The current SFR estimates are derived by fitting the flux of the emission lines, whose luminosities are entirely attributed to star formation, neglecting all the other mechanisms that can produce a ionizing flux, and is taken to be the average during the last 20 Myr. They suffer from aperture effects; total values are computed by rescaling the ones obtained by fitting the optical spectra,  calibrated on the V-band fiber magnitude, to the total V magnitude.
The reliability of the total SFRs were tested comparing the results with independent SFR determinations \citep{2011A&A...526A..45F,2015arXiv150407105P}. 

As discussed in \citet{2014A&A...566A..32F}, emission lines can be measured in our spectra down to a limit of 2 {}\AA, while any emission measurement below this threshold is considered unreliable. This sets a lower detection limit that translates into a specific star formation rate (the SFR per unit of  stellar mass, sSFR) limit of $10^{-12.5}yr^{-1}$. To be conservative, we will consider as star forming only galaxies with 
$sSFR>10^{-12}yr^{-1}$, and we have verified that moving the threshold does not severely affect our results.

To assemble our final sample, we consider all galaxies above the mass completeness limit of $M_\ast>$10$^{9.8}$ M$_{\odot}$, derived as in \citet{2011MNRAS.412..246V}. In clusters, we exclude BCGs and consider all galaxies within 2$R_{200}$. The AGN contribution can be neglected \citep[]{2015MNRAS.450.2749G}. 

We combine the data from all the clusters together, to have better statistics and characterize average properties. 
Our final sample consists of 5065 cluster galaxies and 743 field galaxies (respectively 9242 and 1347 galaxies, once weighted for incompleteness). 

\section{Results}

\subsection{The SFR-M$_*$ relation in clusters and field}
\begin{figure}
 \centering
 \includegraphics[scale=0.45]{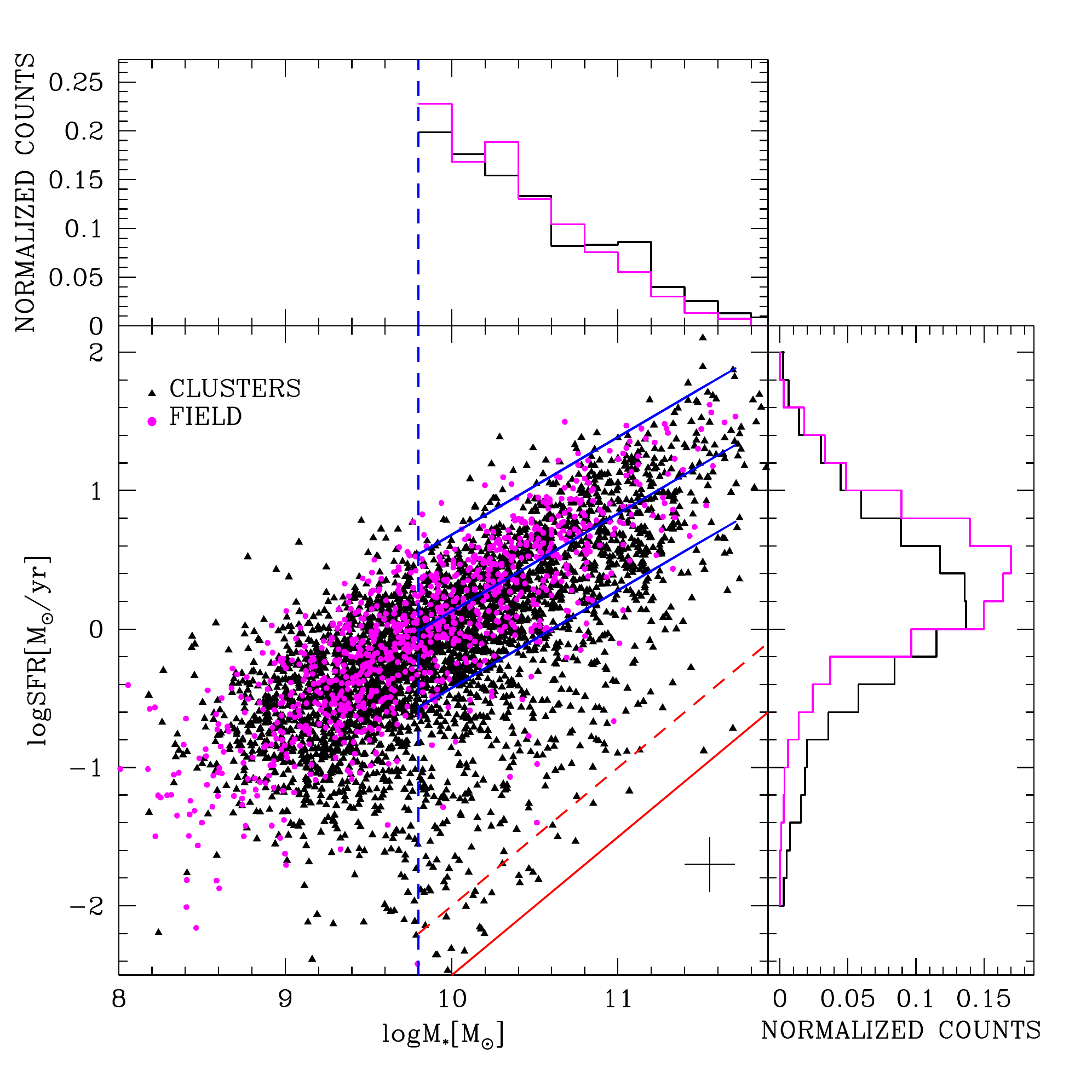}
\caption{Main panel: SFR-M$_*$ relation for cluster (black triangles) and field (magenta circles) galaxies. The blue solid lines indicate the best fit of the field  with 1.5$\sigma$ error, the red solid (dashed) line the $\log sSFR=-12.5(-12)$  limit. The blue vertical dashed line indicates the mass completeness limit. Typical error-bars on SFR and M$_*$ are indicated in the bottom-right corner.  Upper panel: cluster (black) and field (magenta) mass distributions. Right panel: cluster and field SFR distributions above the mass completeness limit. Histograms are normalized to total and weighted for incompleteness. \label{sfrm}}.
 \end{figure}

The main panel of Figure \ref{sfrm} shows the SFR-M$_*$ relation for all SF galaxies in the different environments; the right and top panels show the SFR and M$_*$ distributions, above the mass completeness limit, weighted for incompleteness and normalized to the total. 
 
 \begin{figure*}
 \centering
 \includegraphics[scale=0.5,angle=270]{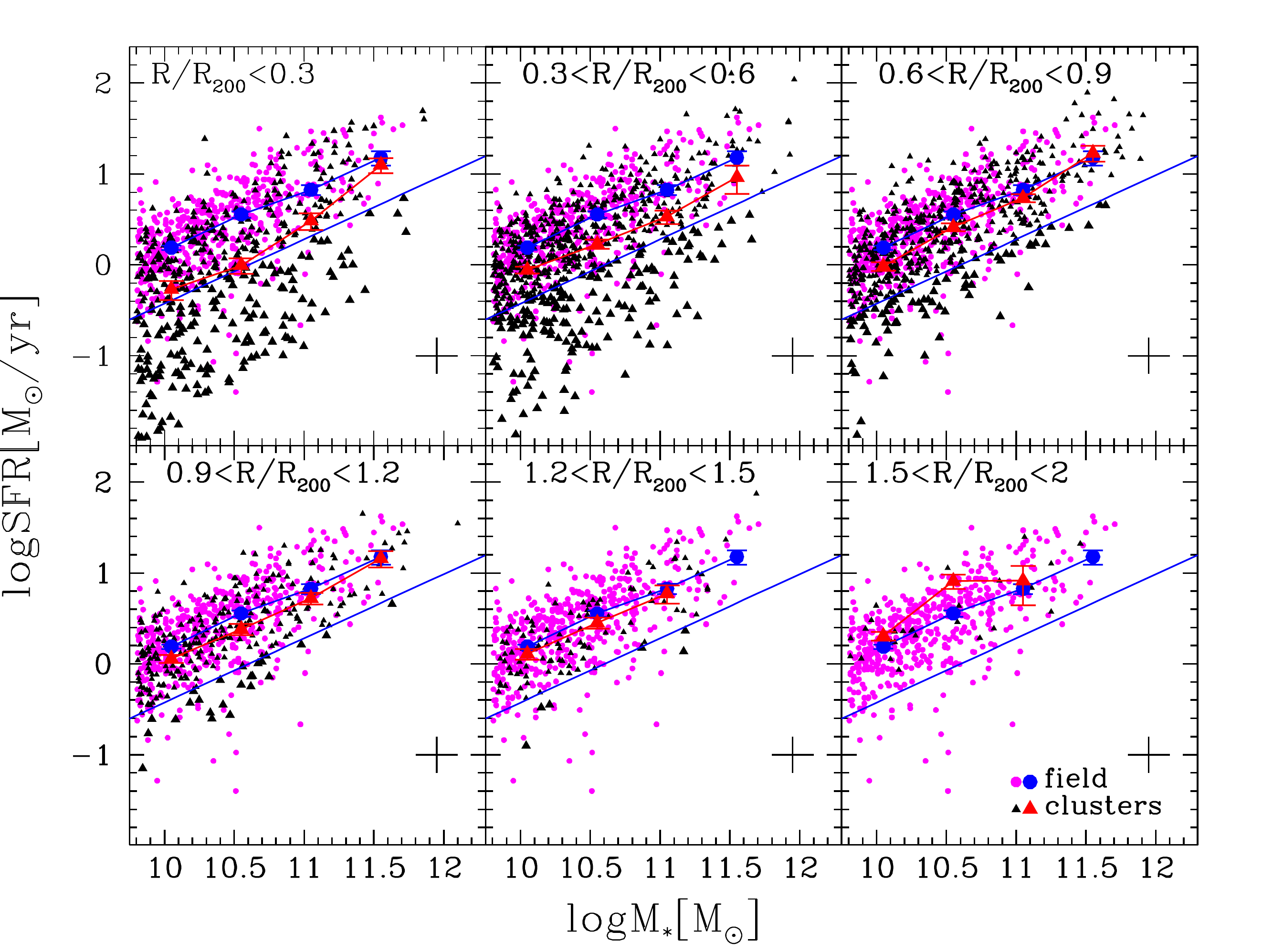}
\caption{SFR-M$_*$ relation above the mass completeness limit in 6 bins of cluster-centric distance, as indicated in the labels. Points and colors are as in Fig. \ref{sfrm}; big triangles indicate transition galaxies. Blue (field) and red (clusters) big symbols are the medians of SF galaxies weighted for incompleteness in different mass bins; error-bars represent the uncertainty on the median. The blue slanting line shows the limit dividing PSF and transition galaxies. Typical error-bars on SFR and M$_*$ are shown in the bottom-right of each panel. \label{sfrm6}}
 \end{figure*} 

Above $M_\ast>$10$^{9.8}$ M$_{\odot}$, the relation for the field can be fitted, accounting for incompleteness, by the equation 
\begin{equation}
\log(SFR)=0.70\times \log (M_*/M_\odot)-6.93
\end{equation}
with a scatter $\sigma \sim0.35$ dex.
This relation agrees with other MSs at similar redshifts \citep[e.g.][]{2007ApJS..173..267S,2007A&A...468...33E} confirming the reliability of our field. 

A change of the SFR-M$_*$ relation with the environment is well visible. 
Even though galaxies in clusters can be as actively star-forming as galaxies in the field, a population with reduced SFRs is evident in the former, while it is much less noticeable in the latter. 

A Kolmogorov-Smirnov (KS) test on the SFR weighted distributions excludes the possibility that the  samples are extracted from the same parent population with a probability $P_{KS}>99.9\%$. The test rejects the null hypothesis also for the two weighted mass distributions,  at a lower level. 
To exclude that the  differences in the SFR distributions are driven by different mass distributions, we performed 100 Monte Carlo simulations extracting randomly a subsample of SF galaxies with the same mass distribution of the field from the cluster sample.
The KS test disproves at a significant level the null hypothesis in 100$\%$ of the cases. 

Hereafter, we further distinguish between {\it purely star-forming} (PSF) and {\it transition} galaxies, the former being within 1.5$\sigma$ from the field fit, the latter lying below -1.5$\sigma$. 
With this cut, transition galaxies represent $\sim24\%$ ($7\%$) of the SF cluster (field) population and 9$\%$ (4$\%$) of the full cluster (field) sample. 

\subsection{Radial trends}

 \begin{figure*}
 \centering
 \includegraphics[scale=0.43]{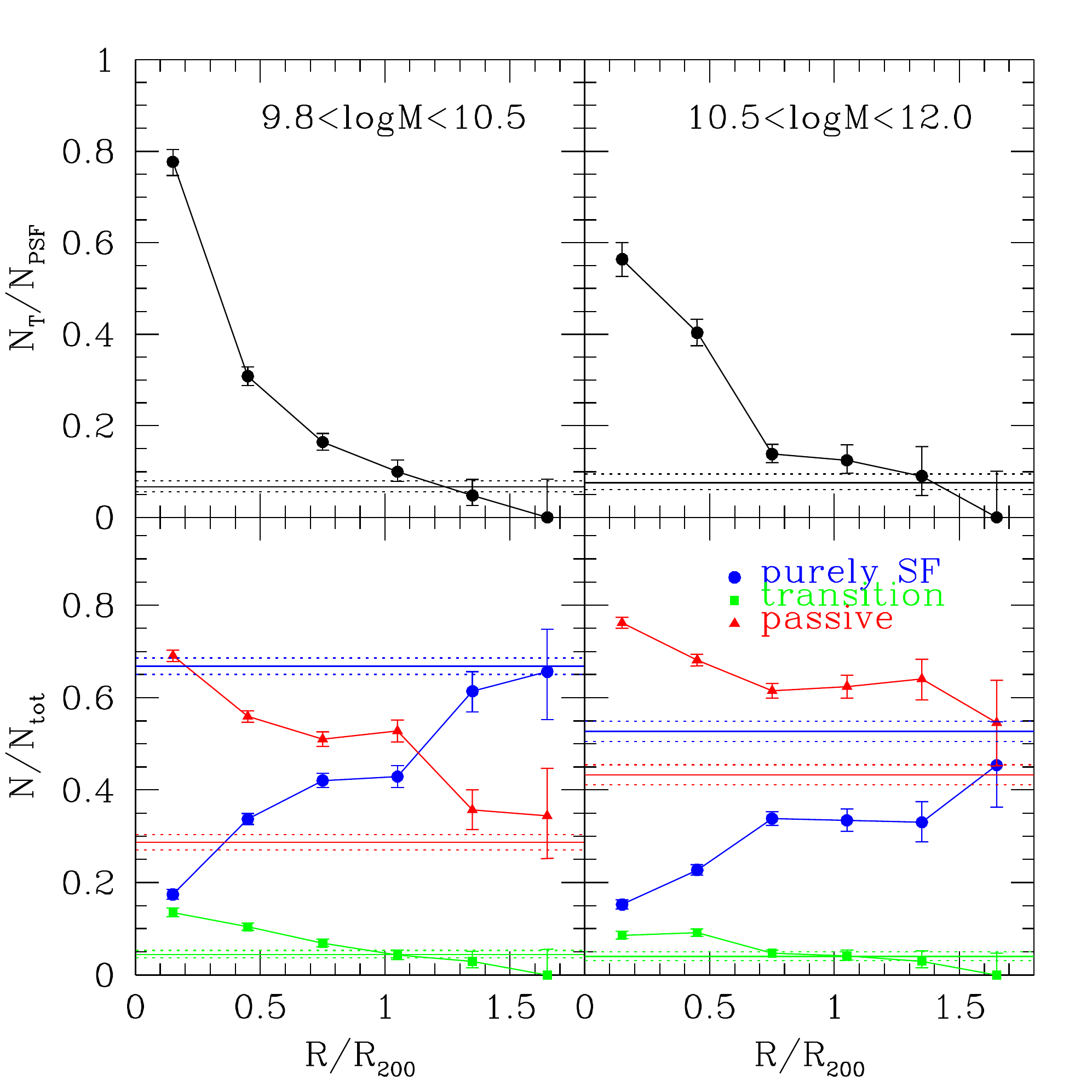}
 \includegraphics[scale=0.43]{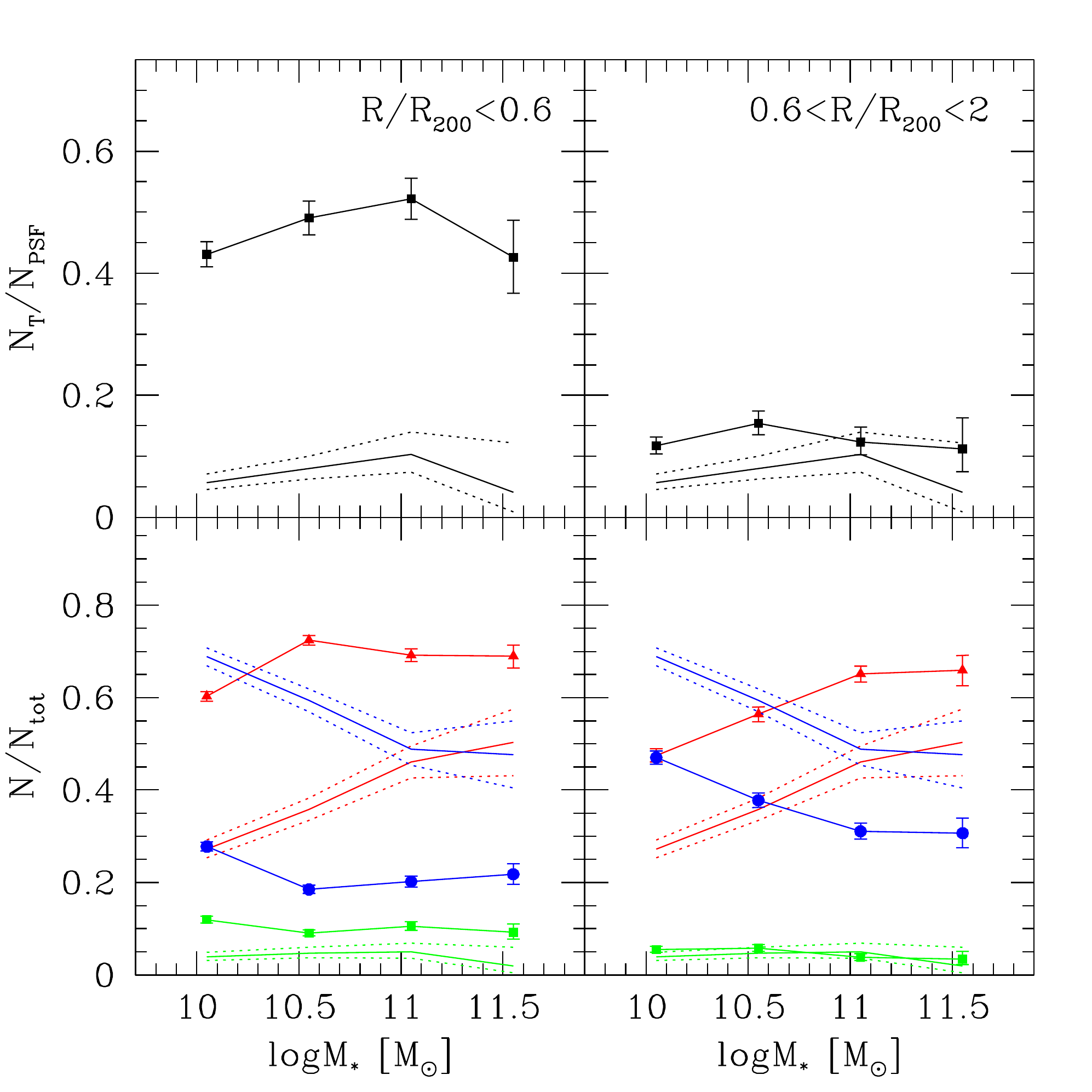}
 \caption{Galaxy fractions. Ratio of transition to PSF galaxies (top panels) and ratio of PSF, passive and transition galaxies to the total (bottom panels) as a function of $R/R_{200}$ in 2 bins of mass (left) and  as a function of $M_*$ in 2 bins of cluster-centric distance (right). 
Points with error bars represent cluster fractions, solid and dotted lines represent field fractions and errors. Errors are binomial \citep{1986ApJ...303..336G}. \label{frac} 
}
 \end{figure*} 
 
We now investigate the spatial distribution of transition galaxies within the clusters (0$<R/R_{200}<$2) and their impact on the SFR-M$_*$ relation.
Figure \ref{sfrm6} shows star-forming galaxies in 6 different bins of projected cluster-centric distance; the field is reported for reference. The median SFRs of all SF galaxies, weighted for incompleteness, have been calculated in 4 mass bins. The quoted uncertainties on the medians are estimated as $1.253\sigma/\sqrt{N}$, where $\sigma$ is the standard deviation about the median and N is the number of galaxies \citep{rider60}. 

Transition galaxies are mainly found within 0.6R$_{200}$, where they represent more than $30\%$ of all SF cluster galaxies and are able to significantly lower the median SFR-$M_*$ relation. In these inner regions, we also notice a lack of PSF galaxies. The frequency of transition galaxies decreases with increasing distance until they almost disappear outside $R_{200}$. 

We note that for $R/R_{200}>1.2$, that corresponds to the radius covered by most of the clusters, the small sample statistics prevents us from drawing solid conclusions, although results do not change if we consider only the 11 clusters that reach 2R$_{200}$.

Both mass and environment play an important role in driving galaxy evolution. 
The median  mass is independent on distance, for PSF, transition and passive galaxies separately (plot not shown), indicating that there is no strong mass segregation. This result suggests that mass and position within the clusters are not strictly related and might play a different role in galaxy quenching. 

We can therefore try to separate the two contributions and understand how the impact of the environment depends on the  mass of the galaxy. The left panel of Fig. \ref{frac} shows the incidence of each sub-population as a function of  cluster-centric distance within two mass bins, while the right panel shows the incidence of each sub-population as a function of mass in two distance bins. Both the mass and radius separation values (respectively $M_*=10^{10.5} M_{\odot}$ and $R/R_{200}=0.6$) used to divide the sample are chosen to have similar numbers of galaxies in each bin. 
The transition to PSF galaxy ratio (top) strongly depends on distance, being $\sim$0.6 at $M_*>10^{10.5}M_{\odot}$ and 0.8 at $10^{9.8}<M_*<10^{10.5}M_{\odot}$ within 0.3$R_{200}$, and rapidly decreases going outward.
The fraction of transition galaxies to the total (bottom) also decreases with distance, but is almost constant with mass, in agreement e.g. with \cite{2011efgt.book...29W, 2015ApJ...798...52V}.

Passive and PSF galaxy fractions strongly depend on both distance and mass. Passive galaxies are the dominant population at all masses inside $R_{200}$ and decrease going outwards, mirrored by PSF galaxies, in agreement with previous studies \cite[e.g.][]{2006MNRAS.366....2W, 2011MNRAS.412..246V}.
Opposite trends are detected in the field, where PSF galaxies represent $\sim$70\% of all galaxies at low masses and 50\% at higher masses, while passive galaxies are $\sim25\%$ at low masses and are nearly as common as PSF galaxies at higher masses. 
In clusters, trends with mass are less pronounced within 0.6$R/R_{200}$, where both PSF and passive fractions are almost constant, but well visible at larger distances, where they resemble what is observed in the field.

Overall, environmental effects seem to dominate within 0.6$R_{200}$: the variation of the relative number of transition and PSF galaxies with distance is the main responsible of the decrease of the median SFR and of the change of the SFR-M$_*$ relation seen in Fig. \ref{sfrm}.

\subsection{Galaxy properties and SFHs}
 \begin{figure*}[!t]
  \centering
 \includegraphics[scale=0.43]{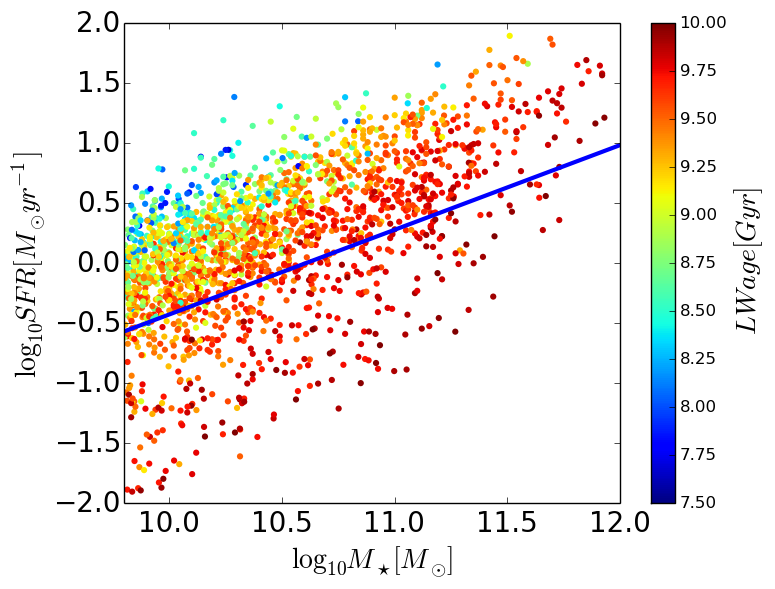}
  \centering
 \includegraphics[scale=0.43]{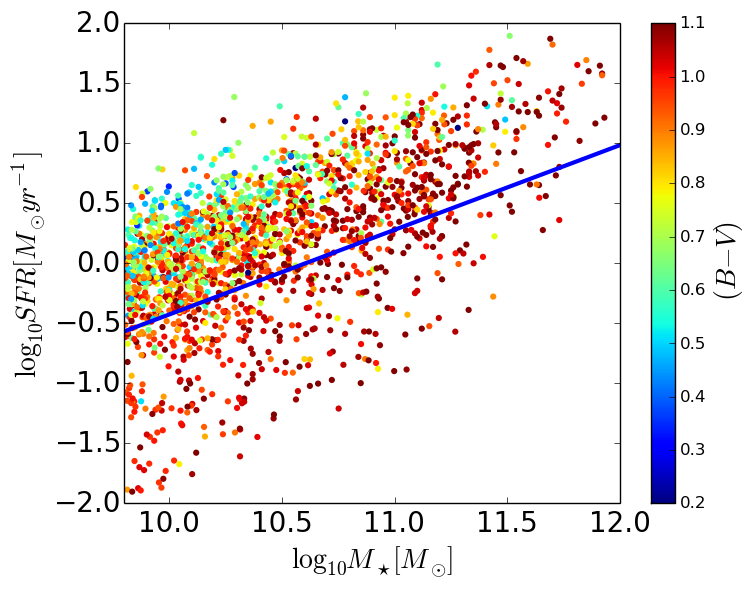}
  \caption{SFR-M$_*$ relation for cluster galaxies. The slanted solid blue line shows the limit dividing PSF and transition galaxies. Color-coded are the luminosity weighted age (left) and (B-V) color (right).  \label{prop} }
 \end{figure*} 
We now investigate whether transition galaxies have different properties from the PSF galaxies. 

Figure \ref{prop} shows how the LWA (left) and (B-V) color (right) change as a function of the position on the SFR-M$_*$ plane for cluster member galaxies.
Moving from left to right and top to bottom, galaxies show redder colors and older LWAs, with the transition population showing the largest values. The median LWA varies from $\sim$5 Gyr for transition galaxies to $\sim$2 Gyr for PSF galaxies, comparable with the field values for all SF galaxies. The same trend is found also for the MWA (plot not shown), but with a less marked gradient.

\begin{figure}
\includegraphics[scale=0.4]{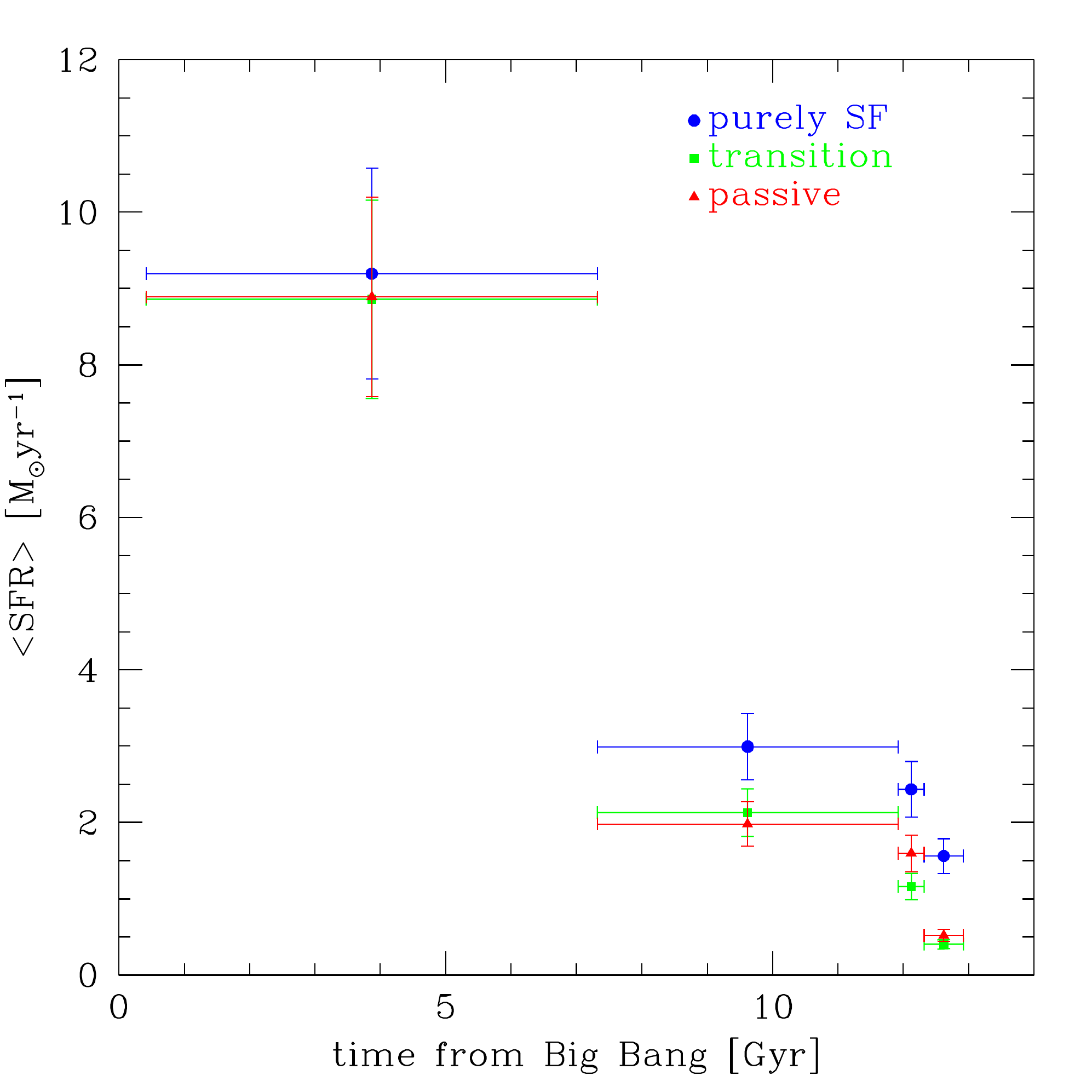}
\caption{Mean SFR as a function of cosmic time for PSF galaxies (blue) and a sample of transition (green) and passive (red) galaxies mass-matched to the PSF population.
Errors on the mean values are obtained using a bootstrap resampling.\label{sfh}}
\end{figure}
We also inspect the SFHs of the different galaxy populations, to trace the evolutionary path of transition galaxies. 
The spectral analysis allows us to derive an estimate of the SFRs at different cosmic times; the initial 12 ages of the SSP spectra were further binned into four intervals with $<t> = 3.9, 9.6, 12, 12.5 $ Gyr and $\Delta t =6.9, 4.6, 4.4, 0.6 $ Gyr \citep[see][]{2015MNRAS.450.2749G}.
Recall that the current SFR, used to separate galaxies, is taken to be the average during the last 20 Myr, thus is slightly different from the first bin of the SFHs. 

To avoid the influence of the different mass distributions, we performed 100 Monte Carlo simulations. 
We randomly extracted a subsample of galaxies with the same  mass distribution of PSF galaxies from the passive and transition samples and computed the mean SFR for each population, in the 4 age intervals. Errors on mean values have been computed as bootstrap standard deviations. Results are shown in Fig. \ref{sfh}.

Overall, the SFH decline gets steeper going from PSF to transition and passive galaxies. Consistently with the analysis of the LWAs, transition galaxies are clearly an evolved population with respect to PSF galaxies, having their SFR suppressed a long time ago (2-5 Gyr).
A more accurate estimation of the quenching timescales is not possible do to the time resolution of our spectrophotometric code.

\section{Discussion and conclusions}
If SF galaxies are affected by environmental mechanisms when they move from the field to groups or clusters, we should see a signature of this transformation that depends on the timescale over which it occurs.

Comparing the SFR-$M_*$ relation of SF galaxies in clusters and in the field and looking at variations as a function of the cluster-centric distance allow us to put constraints on both the range of environments where star formation is quenched and the average timescale over which it happens. 

Rapid quenching processes would leave the SFR-M$_*$ relation unperturbed with respect to the field, moving SF galaxies directly to the red sequence. On the contrary, slow quenching would increase the number of galaxies with reduced SFRs resulting in a different SFR-M$_*$ relation. 
We find a population of low SF galaxies, which is rare in the field, suggesting that for these galaxies the transition from SF to passive occurs on a sufficiently long timescale to let us see them in the process of being quenched. We stress that our analysis cannot identify galaxies quenching on short timescales, that would move quickly from the SF MS to being passive.
We showed that this slow process is confined to galaxies within $R_{200}$ and that only within $0.6R_{200}$ this causes the median SFR-M$_*$ relation to detach from the observed field MS. In addition, in the inner regions galaxy fractions are nearly constant with mass. 

The analysis of the SFHs, together with the variation of the properties of galaxies in the SFR-M$_*$ plane, support this scenario: transition galaxies have older LWAs and redder colors than MS galaxies and show reduced mean SFRs at least in the last $2-5$Gyr.
Stellar mass seems to play a minor role, with the observed trends displaying a much weaker dependence on M$_*$ than on distance.

We cannot exclude that some transition galaxies are passive galaxies returning to the SF population due to accretion of gas-rich satellites or exchange of material during merging, but this channel seems improbable because of the many processes acting to remove gas or suppress the gas reservoir in clusters \citep[see e.g.][]{2015MNRAS.447..969B}.

The results presented here agree with the conclusions of \citet{2010MNRAS.404.1231V} and \citet{2013ApJ...775..126H} at $z=0.1-0.3$, who  found that the SFRs of SF galaxies decline towards the cluster core from the values observed in the field. Also numerical simulations \cite[e.g.][]{2015ApJ...806..101H,2014MNRAS.440.1934T,2013MNRAS.431.2307O} support the slow quenching scenario, finding kinematic segregation between normally star forming galaxies and those with reduced SFRs.

This analysis, together with the work of \citet{2009ApJ...705L..67P} and \citet{2010ApJ...710L...1V}, who  depicted a population of galaxies with reduced SFRs, at $0.4<z\leq 0.8$, suggest that this transition population was already in place at higher redshifts. 

We conclude that, even if cluster galaxies can be as star-forming as field galaxies, there are significant differences in their SFR distributions.
Clusters not only have a pre-existing larger population of passive galaxies, but also show a tail of low star-forming galaxies that is rare in the field. 
This evidence is predicted for strangulation models, in which the diffuse gas halo is rapidly stripped at the passage through the intracluster medium leaving available to the galaxy only its reservoir of molecular gas to form stars \citep[e.g.,][]{2002ApJ...577..651B}. 

A natural step forward will be to further investigate the SFR-M$_*$ relation as a function of local density and the global properties of the clusters, such as $\sigma_{cl}$ and $L_{X}$ and extend the halo mass range covered down to the group scale (Paccagnella et al. in prep.). Such studies will shed further light on the processes causing galaxy quenching and the range of halo masses on which it occurs. 

\section*{Acknowledgments}
We thank the referee for his/her useful remarks.
We acknowledge the financial support from a PRIN-INAF 2014 grant.
AP acknowledges financial support from the Fondazione Ing. Aldo Gini and thanks the Kavli Institute for the Physics and Mathematics of the Universe, the University of Tokyo, for a productive stay during which part of this work  was carried out. 
BV acknowledges the support from the World Premier International Research Center Initiative (WPI), MEXT, Japan and the Kakenhi Grant-in-Aid for Young Scientists (B)(26870140) from the Japan Society for the Promotion of Science (JSPS).


\begin{thebibliography}{}

\bibitem[{Abramson} {\it et al.}\ (2014)]{2014ApJ...785L..36A}
{Abramson}, L.~E. {\it et al.}\  2014, \apjl, 785, L36.

\bibitem[{Bah{\'e}} \& {McCarthy}(2015)]{2015MNRAS.447..969B}
{Bah{\'e}}, Y.~M. and {McCarthy}, I.~G. 2015, \mnras, 447, 969.

\bibitem[{Bai} {\it et al.}\ (2009)]{2009ApJ...693.1840B}
{Bai}, L. {\it et al.}\  2009, \apj, 693, 1840.

\bibitem[{Balogh}, {Navarro} \& {Morris}(2000)]{2000ApJ...540..113B}
{Balogh}, M.~L., {Navarro}, J.~F., and {Morris}, S.~L. 2000, \apj, 540, 113.

\bibitem[{Bekki}, {Couch} \& {Shioya}(2002)]{2002ApJ...577..651B}
{Bekki}, K., {Couch}, W.~J., and {Shioya}, Y. 2002, \apj, 577, 651.

\bibitem[{Bouch{\'e}} {\it et al.}\ (2010)]{2010ApJ...718.1001B}
{Bouch{\'e}}, N. {\it et al.}\  2010, \apj, 718, 1001.

\bibitem[{Brinchmann} {\it et al.}\ (2004)]{2004MNRAS.351.1151B}
{Brinchmann}, J. {\it et al.}\  2004, \mnras, 351, 1151.

\bibitem[{Cava} {\it et al.}\ (2009)]{2009A&A...495..707C}
{Cava}, A. {\it et al.}\  2009, \aap, 495, 707.

\bibitem[{Daddi} {\it et al.}\ (2007)]{2007ApJ...670..156D}
{Daddi}, E. {\it et al.}\  2007, \apj, 670, 156.

\bibitem[{Dressler}(1980)]{1980ApJ...236..351D}
{Dressler}, A. 1980, \apj, 236, 351.

\bibitem[{Elbaz} {\it et al.}\ (2007)]{2007A&A...468...33E}
{Elbaz}, D. {\it et al.}\  2007, \aap, 468, 33.

\bibitem[{Fasano} {\it et al.}\ (2006)]{2006A&A...445..805F}
{Fasano}, G. {\it et al.}\  2006, \aap, 445, 805.

\bibitem[{Fritz} {\it et al.}\ (2007)]{2007A&A...470..137F}
{Fritz}, J. {\it et al.}\  2007, \aap, 470, 137.

\bibitem[{Fritz} {\it et al.}\ (2011)]{2011A&A...526A..45F}
{Fritz}, J. {\it et al.}\  2011b, \aap, 526, A45.

\bibitem[{Fritz} {\it et al.}\ (2014)]{2014A&A...566A..32F}
{Fritz}, J. {\it et al.}\  2014, \aap, 566, A32.

\bibitem[{Gao} {\it et al.}\ (2004)]{2004MNRAS.352L...1G}
{Gao}, L. {\it et al.}\  2004, \mnras, 352, L1.

\bibitem[Gehrels(1986)]{1986ApJ...303..336G} Gehrels, N.\ 1986, \apj, 303, 
336 

\bibitem[{Guglielmo} {\it et al.}\ (2015)]{2015MNRAS.450.2749G}
{Guglielmo}, V. {\it et al.}\  2015, \mnras, 450, 2749.

\bibitem[{Gullieuszik} {\it et al.}\ (2015)]{2015A&A...581A..41G}
{Gullieuszik}, M. {\it et al.}\  2015, \aap, 581, A41.

\bibitem[{Gunn} \& {Gott}(1972)]{1972ApJ...176....1G}
{Gunn}, J.~E. and {Gott}, III, J.~R. 1972, \apj, 176, 1.

\bibitem[{Haines} {\it et al.}\ (2015)]{2015ApJ...806..101H}
{Haines}, C.~P. {\it et al.}\  2015, \apj, 806, 101.

\bibitem[{Haines} {\it et al.}\ (2013)]{2013ApJ...775..126H}
{Haines}, C.~P. {\it et al.}\  2013, \apj, 775, 126.

\bibitem[{Larson}, {Tinsley} \& {Caldwell}(1980)]{1980ApJ...237..692L}
{Larson}, R.~B., {Tinsley}, B.~M., and {Caldwell}, C.~N. 1980, \apj, 237, 692.

\bibitem[{Lilly} {\it et al.}\ (2013)]{2013ApJ...772..119L}
{Lilly}, S.~J. {\it et al.}\  2013, \apj, 772, 119.

\bibitem[{Mihos} \& {Hernquist}(1994)]{1994ApJ...425L..13M}
{Mihos}, J.~C. and {Hernquist}, L. 1994, \apjl, 425, L13.

\bibitem[{Moore} {\it et al.}\ (1996)]{1996Natur.379..613M}
{Moore}, B. {\it et al.}\  1996, \nat, 379, 613.

\bibitem[{Moretti} {\it et al.}\ (2014)]{2014A&A...564A.138M}
{Moretti}, A. {\it et al.}\  2014, \aap, 564, A138.

\bibitem[{Noeske} {\it et al.}\ (2007)]{2007ApJ...660L..43N}
{Noeske}, K.~G. {\it et al.}\  2007, \apjl, 660, L43.

\bibitem[{Oman} {\it et al.}\ (2013)]{2013MNRAS.431.2307O}
{Oman}, K.~A {\it et al.}\  2013, \mnras, 431, 2307.

\bibitem[{Patel} {\it et al.}\ (2009)]
{2009ApJ...705L..67P}
{Patel}, S.~G. {\it et al.}\  2009, \apjl, 705, L67.

\bibitem[{Peng} {\it et al.}\ (2010)]{2010ApJ...721..193P}
{Peng}, Y.-j. {\it et al.}\  2010, \apj, 721, 193.

\bibitem[{Poggianti} {\it et al.}\ (2009)]{2009ApJ...697L.137P}
{Poggianti}, B.~M. {\it et al.}\  2009, \apjl, 697, L137.

\bibitem[{Poggianti} {\it et al.}\ (2015)]{2015arXiv150407105P}
{Poggianti}, B.~M. {\it et al.}\  2015, ArXiv e-prints.

\bibitem[{Poggianti} {\it et al.}\ (1999)]{1999ApJ...518..576P}
{Poggianti}, B.~M. {\it et al.}\  1999, \apj, 518, 576.

\bibitem[{Poggianti} {\it et al.}\ (2006)]{2006ApJ...642..188P}
{Poggianti}, B.~M. {\it et al.}\  2006, \apj, 642, 188.

\bibitem[{Rider}\ (1960)] {rider60}
{Rider}, P. R.\ 1960, JASA, 55, 148.

\bibitem[{Salim} {\it et al.}\ (2007)]{2007ApJS..173..267S}
{Salim}, S. {\it et al.}\  2007, \apjs, 173, 267.

\bibitem[{Salpeter}(1955)]{1955ApJ...121..161S}
{Salpeter}, E.~E. 1955, \apj, 121, 161.

\bibitem[{Taranu} {\it et al.}\ (2014)]{2014MNRAS.440.1934T}
{Taranu}, D.~S. {\it et al.}\  2014, \mnras, 440, 1934.


\bibitem[{Tyler}, {Bai} \& {Rieke}(2014)]{2014ApJ...794...31T}
{Tyler}, K.~D., {Bai}, L., and {Rieke}, G.~H. 2014, \apj, 794, 31.

\bibitem[{Tyler}, {Rieke} \& {Bai}(2013)]{2013ApJ...773...86T}
{Tyler}, K.~D., {Rieke}, G.~H., and {Bai}, L. 2013, \apj, 773, 86.

\bibitem[{van den Bosch} {\it et al.}\ (2008)]{2008arXiv0805.0002V}
{van den Bosch}, F.~C. {\it et al.}\  2008, ArXiv e-prints.

\bibitem[{von der Linden} {\it et al.}\ (2010)]{2010MNRAS.404.1231V}
{von der Linden}, A. {\it et al.}\  2010, \mnras, 404, 1231.

\bibitem[{Vulcani} {\it et al.}\ (2010)]{2010ApJ...710L...1V}
{Vulcani}, B. {\it et al.}\  2010, \apjl, 710, L1.

\bibitem[{Vulcani} {\it et al.}\ (2011)]{2011MNRAS.412..246V}
{Vulcani}, B. {\it et al.}\  2011, \mnras, 412, 246.

\bibitem[{Vulcani} {\it et al.}\ (2015)]{2015ApJ...798...52V}
{Vulcani}, B. {\it et al.}\  2015, \apj, 798, 52.

\bibitem[{Weinmann} {\it et al.}\ (2006)]{2006MNRAS.366....2W}
{Weinmann}, S.~M. {\it et al.}\  2006, \mnras, 366, 2.

\bibitem[{Weinmann}, {van den Bosch} \& {Pasquali}(2011)]{2011efgt.book...29W}
{Weinmann}, S.~M., {van den Bosch}, F.~C., and {Pasquali}, A. 2011.
\newblock { {The Dependence of Low Redshift Galaxy Properties on Environment}},
  29.

\bibitem[{Wijesinghe} {\it et al.}\ (2012)]{2012MNRAS.423.3679W}
{Wijesinghe}, D.~B. {\it et al.}\  2012, \mnras, 423, 3679.

\end{thebibliography}
\end{document}